\documentclass[preprint]{elsarticle}  
\usepackage{latexsym}
\usepackage{amssymb}
\usepackage{amsmath}
\usepackage{graphicx}
\usepackage{multirow}

\usepackage{graphicx}

\newcommand{\lab}[1]{\label{#1}}
\newcommand{\ba}{\begin{eqnarray}}
\newcommand{\ea}{\end{eqnarray}}
\newcommand{\beqs}{\begin{eqnarray}}
\newcommand{\eeqs}{\end{eqnarray}}

\newcommand{\dd}{\Delta}

\begin{document}

\begin{center}
{\Large {\bf
Anomalies in the differential cross sections at 13 TeV. 
}}
\end{center}

\begin{center}
{\bf O.V. Selyugin} 
\end{center}


\begin{center}
 { BLTPh, JINR, Dubna, RF}
\end{center}

\begin{center}
\end{center}



\section*{Abstract}
{
 Analysis of  
    differential cross sections of  the TOTEM Collaboration data, carried out
  without model assumptions,
  shows the existence of  new effects in the behavior of the
    hadron scattering amplitude at  a small momentum transfer at
    a high confidence level.
    The quantitative description of the data in the framework of the
    high energy generalized structure (HEGS)
    model supports such a phenomenon that
    can be associated with the specific properties of the  hadron potential
  at large distances.
    It is shown that the value of $\rho(s,t)$ at $\sqrt{s}=13$ TeV and  small $t$
    exceeded $0.1$.
}


\section{Introduction}
\label{sec:intro}
    The unique experiment carried out by
     the  TOTEM Collaboration  at LHC at 13 TeV gave  excellent experimental data
     on the elastic proton-proton scattering in a wide region of transfer momenta \cite{T66,T67}.
      It is  especially necessary to note the experimental data obtained at small momentum transfer
     in the Coulomb-hadron interference region. The experiment reaches  very small
     $t = 8 \ 10^{-4}$ GeV$^2$  with small $\Delta t$, which gave a large number of
     experimental points in a sufficiently small region of momentum transfer.
     This allows one to carry out careful analysis of the experimental data to explore
     some properties of the hadron elastic scattering.

   There are two sets of  data - at small momentum transfer \cite{T66}
   and at middle and large momentum transfer \cite{T67}.
   They overlap in some region of the momentum transfer which supply practically the same
   normalization of both sets of the  differential cross section
   of  elastic proton-proton scattering.

    A  research of the structure of the elastic hadron scattering amplitude
   at superhigh energies and small momentum transfer - $t$
  can give a connection   between
  the  experimental knowledge and  the basic asymptotic theorems
    based on  first principles \cite{akm,fp,royt}.
   It gives   information about the  hadron interaction
   at large distances where the perturbative QCD does not work \cite{Drem,PetOkor-18},
   and a new theory as, for example, instanton or string theories
    must be developed.

    There is a very important  characteristic of the elastic scattering amplitude such as
    the ratio of the real part to imaginary part of the scattering amplitude - $\rho(s,t)$.
    It is tightly connected with the integral and differential dispersion relations.
    Of course, especially after different results obtained by the UA4  and UA4/2 Collaborations at SPPS
     physicists understand that $\rho(s,t=0)$ is not simple experimental value but heavily
    dependent on  theoretical assumptions about the momentum dependence of the elastic scattering
    amplitude. Our analysis of both experimental data obtained by the UA4  and UA4/2 Collaborations,
    shows a small difference  value of  $\rho(s,t=0)$   between  both the experiments if the nonlinear
    behaviour of the elastic scattering amplitude is taken into account \cite{Sel-UA42}.
      Hence, this is  not a experimental problem
   but a theoretical one  \cite{CS-PRL09}.

           Usually,  a small region of  $t$ is taken into account
   for extraction of the sizes of $\sigma_{tot}$ and $\rho(t=0)$ (for example \cite{T66,Protvino19}).
  However, a form of the scattering amplitude assumed for small $t$ and satisfying the existing experimental data
  at small $t$, can essentially be different from experimental data at large momentum transfer.
   One should take into account the analysis of  the differential cross section at 13 TeV where the diffraction minimum impacts the form of $d\sigma/dt$  already at $t = -0.35$ GeV$^2$.

%
%
%
%
%
%
%
%

      \section{Thin effects in the differential cross sections at $\sqrt{s}=13$ TeV}

   The extraction of  values of the basic parameters of the elastic hadron interaction
   requires some model that can describe all experimental data
   at the quantitative  level with minimum free parameters.
   Now  many groups of researchers have presented some physical models satisfying
   more or less  these   requirements. This is  especially   related with the HEGS
   (High Energy Generalized Structure) model \cite{HEGS0,HEGS1}.
   As it takes into account  two form factors (electromagnetic and gravitomagnetic),
   which are calculated from the GPDs function of nucleons, it has a minimum of free
   parameters and gives a quantitative description of the exiting experimental data
   in a wide energy region and momentum transfer.
   One of the specific properties of our analysis is that in the fitting procedure we take into account
   only statistical errors. The systematic errors are taken as an additional coefficient
   which changes the normalization of one set of  experimental data. In this case,
   the space for  theoretical functions decreases essentially but can lead to an increase in the
   whole $\chi^2$  \cite{Orava-Sel}.

  However,  let us
  carried out analysis of  $d\sigma/dt$ of the TOTEM Collaboration data,
  without model assumptions to catch out some possible think effects like some periodical
  structure
   in the behavior of the
    hadron scattering amplitude at  a small momentum transfer
   \cite{Osc-13}.

 For  model free analysis, let us  use the method of
 comparison of two statistically independent  choices,
  for example \cite{hud}.
  Such method does not require  knowledge of the form of the additional   periodic
  part of the scattering amplitude.
   If we have two statistically independent choices
  $x^{'}_{n_1}$  and $x^{"}_{n_2}$
  of values of the quantity  $X$  distributed around
  a definite value of $A$ with the standard error  equal to $1$,
  We can find the difference between
  $x^{'}_{n_1}$  and $x^{"}_{n_2}$. For that, we can compare
  the arithmetic mean of these choices
$$ \dd X = (x^{'}_1 + x^{'}_2 + ... x^{'}_{n1})/n_1  -
        (x^{"}_1 + x^{"}_2 + ... x^{"}_{n2})/n_2  =
     \overline{x^{'}_{n_1}} - \overline{x^{"}_{n_2}}.   $$
  The standard deviation for this case will be
$   \delta_{\overline{x}} = [1/n_1 +1/n_2]^{1/2} $.
 And if $\dd X / \delta_{\overline{x}}$ is larger than $3$, we can say
 that the difference between these two choices has  the
  $99\%$ probability .

The deviations $\Delta R_i$ of experimental data from these
 theoretical cross
 sections we will be measured in units of experimental error for
 an appropriate point
\ba
\Delta R_ {i} = [(d\sigma/dt_{i})^{exp}
  -  (d\sigma/dt_{i})^{th}] / \delta_{i}^{exp},
\ea
 where $\delta_{i}^{exp} $ is an experimental error.
 To take this effect into account, we  break
 the whole studied interval of momentum transfer into $k$  equal pieces
 $k \delta t = (t_{max} -t_{min}) $,
 and then
  sum  $\Delta R_{i} $
 separately over even and odd pieces.
 Thus, we receive two sums $S^{up} $ and $S^{dn} $ for
  $n_1$ even and $n_2$ odd intervals. At this
  $n_1 + n_2 = k$ and $|n_1 - n_2| = 0$ or $1$
  \ba
   S^{up} = \sum_ {j=1}^{n_1} (\sum^{N}_{i}
            \Delta R_{i})|_{\delta q (2j-1) < q_i \leq \delta q(2j) }, \ \ \ 
S^{dn} = \sum_{j=1}^{n_2} (\sum^{N}_{i=1} \Delta R_{i})
|_{\delta q (2j) < q_i \leq  \delta q (2j + 1) }.
\label{sdn}
\ea
 In the case of some difference of experimental data from
 the theoretical behavior,
 expected by us, or incorrectly determined parameters, these two sums
 will deviate from zero; but their sizes should coincide within
 experimental errors.
 However, this will be so in the case if experimental data have no any
 periodic structure or a sharp effect coincides with one interval.
 We assume that such a periodic structure is available
 and its period coincides with the chosen interval $2 \delta t$. In this
  case, the sum
 $S^{up} $ will contain, say, all positive half-cycles;
 and the sum $S^{dn} $, all negative half-cycles.
 The difference between $S^{up} $ and
 $S^{dn}$ will show the magnitude of an additional effect summed
 over the whole researched domain.

 The method  does not require  exact representation of the
 periodical part of the scattering amplitude, and now let us apply it to  new LHC data
 of the TOTEM Collaboration at $13$ TeV. 
 The region of momentum transfer  examined up to $-t < 0.4$ GeV$^2$
 includes the Coulomb-hadron  interference range.
%
Of course, it is necessary  to choose the true interval $\delta t$ to obtain the maximum
of the difference between the sums   $S^{up} $ and $S^{dn} $.
To evaluate the size of a possible effect, one should examine the difference
 of the arithmetic mean values
   $\dd S$ and the corresponding dispersion - $\delta S$ \cite{hud}
\ba
\overline{ \dd S} = \overline{S^{up}} - \overline{S^{dn}}; \ \ \overline \delta S = (1/[1/n_1 + 1/n_2]^{1/2})/N.
\ea

 Let us calculate the sum of $\overline{ \dd S}$  and its
 arithmetic mean chosen  in the most appropriate interval $\delta t$   \cite{Osc-13}
\ba
\overline {\Delta S} =285/325 = 0.877 \pm 0.028.
\ea
  Obviously, this is shows the existence of  some periodical structure 
    at a high   confidence level.

   Now let us try to find the form of such an additional periodical contribution
     to the basic elastic scattering amplitude.
     As a basis, take our high energy generalized structure (HEGS) model \cite{HEGS0,HEGS1} which quantitatively  describes, with only a few parameters, the   differential cross section of $pp$ and $p\bar{p}$
    from $\sqrt{s} =9 $ GeV up to $13$ TeV, includes the Coulomb-hadron interference region and the high-$|t|$ region  up to $|t|=15$ GeV$^2$
    and quantitatively well describes the energy dependence of the form of the diffraction minimum \cite{HEGS-min}.
     However, to avoid  possible problems
   connected with the low-energy region, we consider here only the asymptotic variant of the model \cite{HEGSh}.

    The total elastic amplitude in general receives five helicity  contributions, but at
   high energy it is enough to write it as $F(s,t) =
  F^{h}(s,t)+F^{\rm em}(s,t) e^{\varphi(s,t)} $\,, where
 $F^{h}(s,t) $ comes from the strong interactions,
 $F^{\rm em}(s,t) $ from the electromagnetic interactions and
 $\varphi(s,t) $
 is the interference phase factor between the electromagnetic and strong
 interactions \cite{selmp1}.
    The Born term of the elastic hadron amplitude at large energy can be written as
    a sum of two pomeron and  odderon contributions.
 All terms are supposed to have the same intercept  $\alpha_0=1+\epsilon_0 = 1.11$, and the pomeron
 slope is fixed at $\alpha'= 0.24$ GeV$^{-2}$.
  The model takes into account  two hadron form factors $F_1(t)$ and $A(t)$, which correspond to  the charge and matter
  distributions \cite{GPD-PRD14}. Both form factors are calculated  as the first and second moments of  the same Generalized Parton Distributions (GPDs).

 As a probe for  the oscillatory function take \cite{SelOsc-PL}
\vspace{-0.1 cm}
\ba
 f_{osc}(t)=h_{osc} (i+\rho_{osc}) J_{1}(\tau))/\tau;  \ \tau = \pi \ (\phi_{0}-t)/t_{0},
\ea
here $J_{1}(\tau)$ is the Bessel function of the first order.
 This form has only a few additional fitting parameters and allows one to represent
 a wide range of  possible oscillation functions.

  After the fitting procedure, we obtain  $\chi^2/dof =1.24$ (remember that we used only statistical errors).
 One should note that the last points of the second set above $-t=2.8$ GeV$^2$  show
 an essentially different slope, and we removed them. The total number of  experimental points
 of both sets of the TOTEM Collaboration equals $415$. If we remove the oscillatory function, then
  $\chi^2/dof =2.7$, so an increase is more than two times.  If we make a new fit without $f_{osc}$,
 then  $\chi^2/dof =2.5$ decreases but remains large.

 To see the oscillations in the differential cross sections, let us
  determine two values - one is pure by theoretical and the  other with the experimental data
\ba
 R\Delta_{th} &=& [d\sigma/dt_{th0+osc} - d\sigma/dt_{th0}]/d\sigma/dt_{th0}, \nonumber \\
 R\Delta_{Exp}&=& [d\sigma/dt_{Exp} - d\sigma/dt_{th0}]/d\sigma/dt_{th0}.
\ea
  The corresponding values calculated from the fit of two sets of the TOTEM data at $13$ TeV are presented in Fig.1.

   However, the additional normalization coefficient reaches a sufficiently large value,
     about $13\%$. It can be in a large momentum transfer region but is very unusual for a small
     momentum transfer. However, both sets of experimental data (small and large region of $t$)
     overlap in some region and, hence,  affect  each other's normalization.
     It is to be noted, that the size of the normalization coefficient does not impact the size
     and properties of the oscillation term. We have examined  many different variants of our
     model (including large and unity normalization coefficient) , but the parameters of
     the oscillation term have small variations.



\begin{figure}
\vspace{-2.cm}
\begin{center}
 \includegraphics[width=0.7\textwidth]{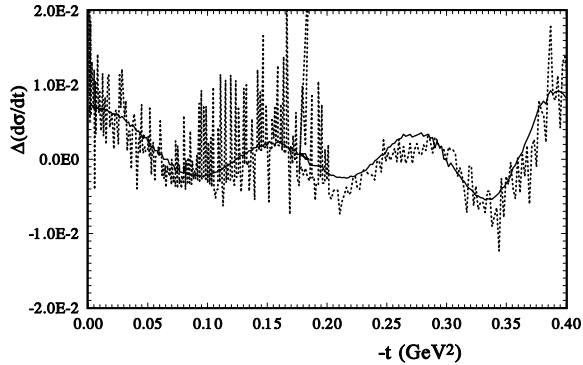}
 \end{center}
\vspace{1.cm}
\caption{ $R\Delta{th} $ of eq.(6a) (the hard line) and $R\Delta{ex} $
  eq.(6b) (the tiny line)  at  $\sqrt(s)=13 $ TeV.
  }
\label{Fig_1}
\end{figure}

In the  work \cite{fd-13}, the analysis of both sets of the TOTEM data at 13 TeV
     is carried out with  additional normalization equal to unity and taking into account
     only  statistical errors in experimental data.
  The Born scattering amplitude has  four free parameters (the constants $C_{i}$) at high energy:
two for the two pomeron amplitudes  and two for the odderon.
The real part of the hadronic elastic scattering amplitude is determined
   through the complexification $\hat{s}=s \ exp(-i \pi/2)$ to satisfy the dispersion relations.

  Now let us put the additional normalization coefficient to unity and continue to take into account
  in our fitting procedure only statistical errors. Of course, we obtain an enormously huge
   $\sum \chi^2$. The new fit changes the basic parameters of the Pomeron and Odderon Born terms
  but does not lead to a reasonable  size of  $\chi^2$.
  We find that the main part of    $\sum \chi^2$ comes from the region of a very small momentum transfer.
  It requires the introduction of a new term which can help to describe the CNI region of $t$.
   This kind of term can be taken in  different forms. In the present paper, we examined
  \ba
 F_{d}(t)=h_{d} (i+\rho_{d}) e^{-B_{d} |t|^{\kappa} \log{\hat{s} }},
 \label{fd-exp}
\ea
 where $ G_{el}^2 $ is the squared electromagnetic form factor of the proton.
  For simplicity, in a further fitting procedure  the constant $\rho_{osc}$  and the phase $\phi_{0}$
  of the oscillatory term are taken equal to zero. Hence, the  oscillatory term  depends only on two parameters -
  $h_{osc}$ and $t_{0}$ period of  oscillation. Also, to reduce the number of fitting parameters
  the correction to the main slope, determined by $\pi$-meson loop \cite{RevJenk}, is taken in a simple form, we obtain the slope as
   \ba
 B(t)= \alpha^{'} \log{\hat{s}} ( 1- t e^{ B_{ad} t}).
 \label{slope-s}
\ea

\begin{figure}
%
\begin{center} 
\includegraphics[width=0.45\textwidth]{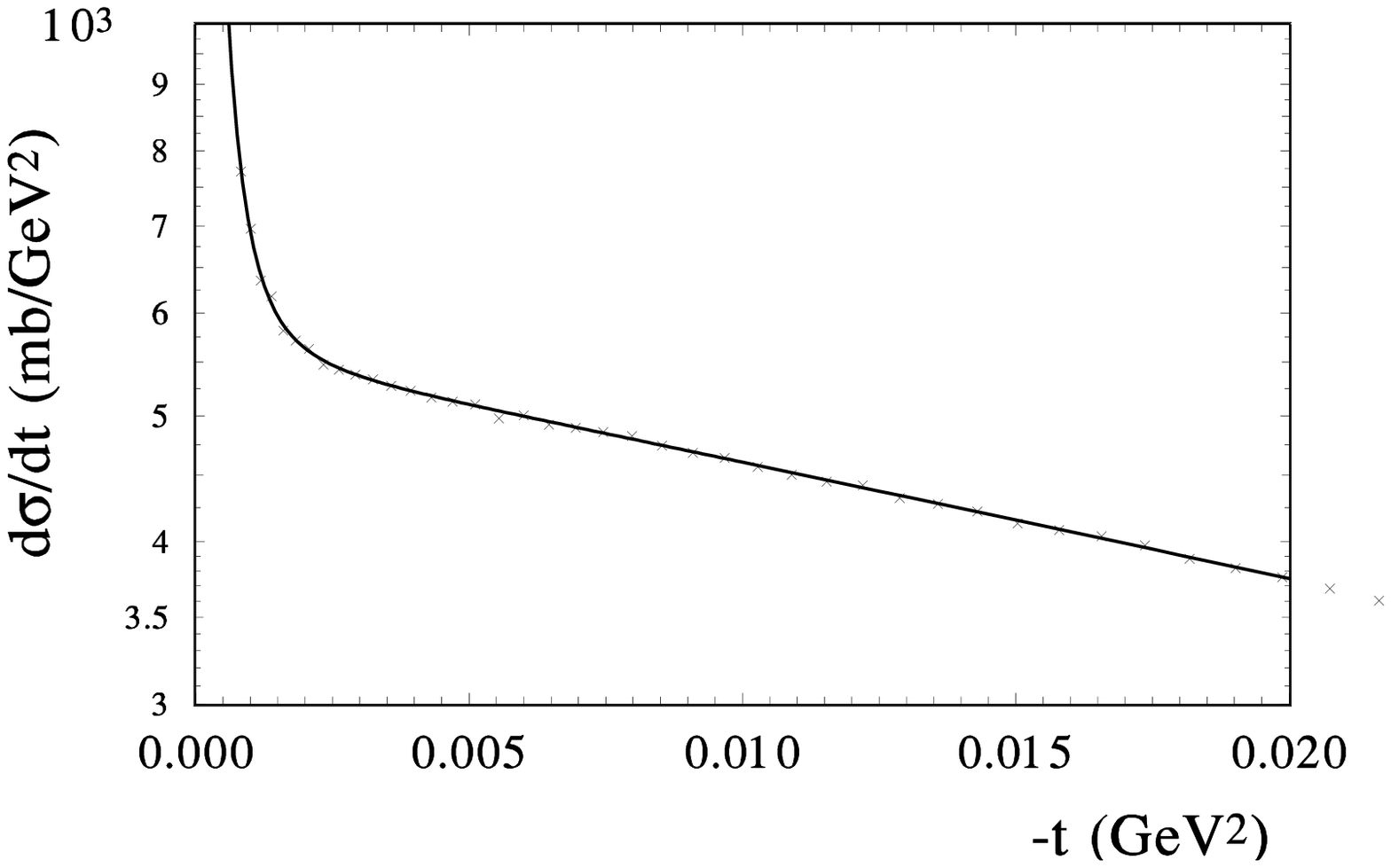}
 \includegraphics[width=0.45\textwidth]{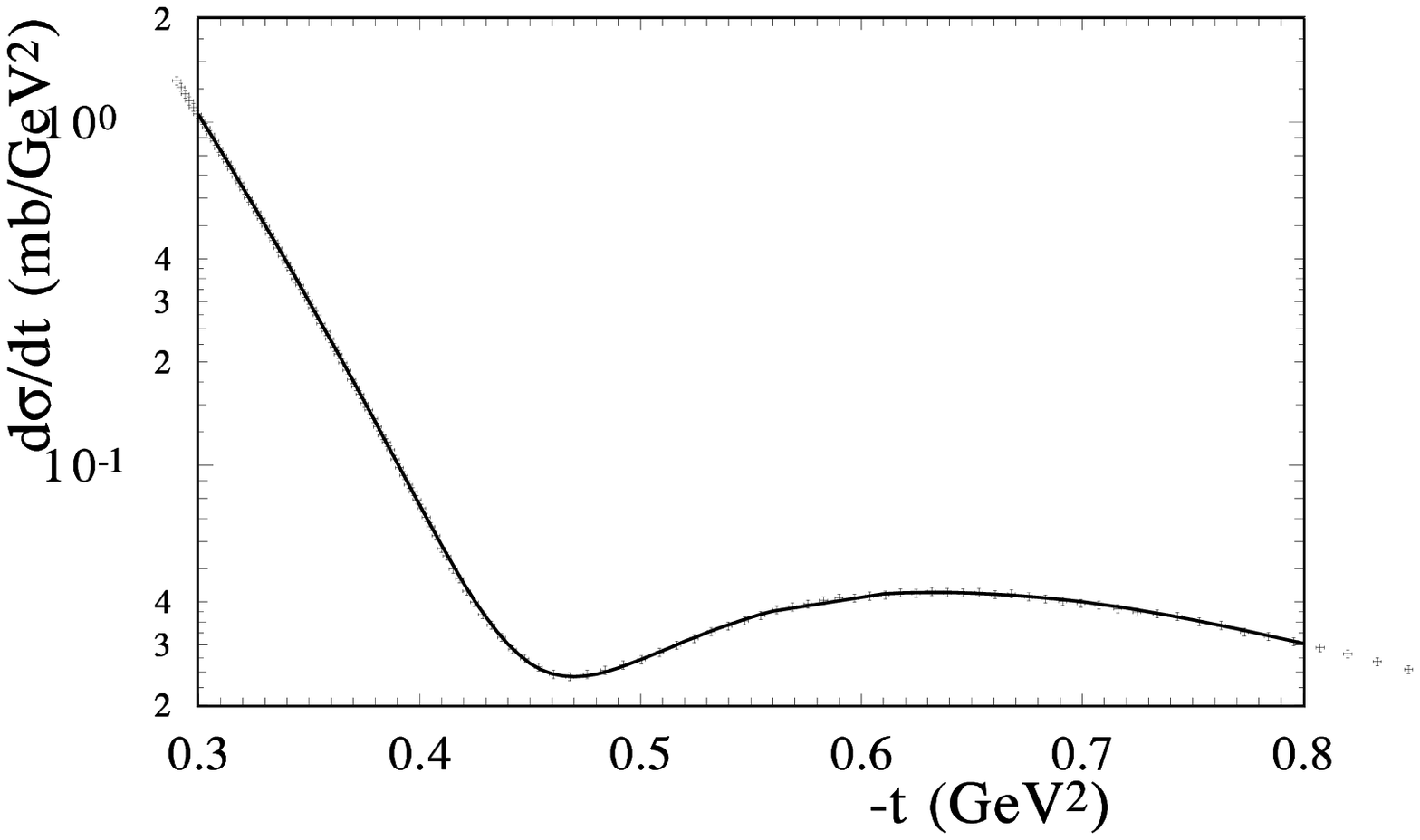}
\end{center}
\vspace{0.5cm}
\caption{The differential cross sections are calculated in the framework of the HEGS model
 with fixed additional normalization by $1.0$ and with additional term 
   a) [left] the magnification of the region of the small momentum transfer of a);
  b) [right] the magnification  of the region of the diffraction minimum.
  }
\label{Fig_2}
\end{figure}

  The fit of both sets of the TOTEM data simultaneously with taking into account only statistical errors, with additional normalization equal to unity and with the additional term,  eq.(7), gives
  a very reasonable $\chi^2 = 551/425=1.29$. The results are presented for
   zoom of the region of small $t$ in Fig.2a, and zoom of the region
  of the diffraction minimum in Fig.2b.

   The parameters of the additional term are well defined  \\
   $ h_{d}=1.7 \pm 0.01;  \ \ \ \rho_{d}=-0.45 \pm 0.06; $ \ \
   $ B_{d} = 0.616 \pm 0.026;  \ \ \  \kappa=1.119 \pm 0.024. $

 To check up the impact of the form of the CNI  phase - $\varphi (t)$, we made our calculations
 with  the original Bethe phase $\varphi  = -( Ln(B_{sl}/2. \ t)+0.577)$ as well.
 We found that $\sum \chi^2$
 changes by less than $0.2\%$ and practically does not impact  the parameters $F_d (t)$.
  Hence, our model calculations show  two possibilities in the quantitative
  description of the two sets of the TOTEM data.
  One - takes into account an additional normalization coefficient, which
  has a minimum size of about $13\%$ ; the other - the introduction of a new anomalous term
  of the scattering amplitude, which has a very large slope and gives the main contributions
  to the Coulomb-nuclear interference region.

    Of course, there are some other ways to obtain good descriptions of the
    new experimental data of the TOTEM Collaboration. One is  to use  a model
    with an essentially increasing  number of the fitting parameters and many different
    parts of the scattering amplitude. Another is to use  a polynomial model
    with many free parameters.
    In both cases, the physical value of such a description is doubtful.

\section{Size of $\rho(\sqrt{s}=13 \ TeV,t)$ }

    There is a large discussion about the value of the $\rho(s,t=0)$ - the ratio of the real to imaginary part
    of the elastic scattering amplitude at $\sqrt{s}=13$ TeV. If the TOTEM Collaboration givs the size of
     $\rho(s,t=0)=0.09 \pm 0.01$ \cite{T66} using own simple phenomenological analysis,
      other researchers obtained the value $\rho(s,t=0)$ some above that value using the model description
     of the differential cross sections in a wide region of momentum transfer.
     For example in \cite{Panch18} it is noted that "... the value of $\rho$ would be higher than the TOTEM value for $\rho$
     found under the hypothesis that the real part of the elastic nuclear amplitude is devoid of such a zero
     in the CNI region."

       There is some simple method to obtain the value of $\rho(s,t=t_{CN})$ at  one point $t_{CN}$
       without any assumptions of the momentum transfer behavior of the real part of the elastic scattering amplitude
       and check up some model assumptions.
 Let us consider from this point of view  experimental data on
 nucleon-nucleon elastic scattering, being available in the range of
small transfers of a pulse.

   The differential cross sections measured experimentally
 are described by the squared scattering amplitude
\ba
d\sigma /dt &=& \pi \ (F^2_C (t)+ (1 + \rho^{2} (s,t)) \ Im F^2_N(s,t)
                                                            \nonumber \\
  & & \mp 2 (\rho (s,t) +\alpha \varphi )) \ F_C (t) Im F_N(s,t)).   \label{ds2}
\ea
where $F_{C} = \mp 2 \alpha G^{2}/|t|$ is the Coulomb amplitude;
$\alpha$ is the fine-structure constant  and $G^{2}(t)$ is  the  proton
electromagnetic form factor squared;
$Re\ F_{N}(s,t)$ and $ Im\ F_{N}(s,t)$ are the real and
imaginary parts of the nuclear amplitude;
$\rho(s,t) = Re\ F(s,t) / Im\ F(s,t)$.
Just this formula is used to fit   experimental  data
determined by the Coulomb and hadron amplitudes and the Coulomb-hadron
phase to obtain the value of $\rho(s,t)$.

  From equation (\ref{ds2}) one can obtain an equation for
 the real (or imaginary) part of the scattering amplitude
  or for $\rho$
   for every experimental point - $i$ if we take the ordinary
  exponential form for the imaginary (or real) part of the scattering
  amplitude \cite{selyf}
\ba
  Re F_N(s,t)=
 ( ReF_c(s,t) +\sqrt{\frac{1}{\pi}\frac{d\sigma}{dt}n - (ImF_c+ImF_N)^2}.
                                                                 \lab{rsq}
\ea
 As the imaginary part of the scattering amplitude is defined by
\ba
  ImF_N(s,t) = \frac{\sigma_{tot}}{4 \pi *0.389} exp(B t /2 ),
\ea
 it is evident from  (\ref{rsq}) that the real part depends on
 $n, \sigma_{tot}, B$. It is clear that if the differential cross sections
 have a special dependence,
  this will manifest itself most strongly in the calculated
 real part of the hadron scattering amplitude.
\begin{figure}
%
\begin{center}
\includegraphics[width=0.7\textwidth]{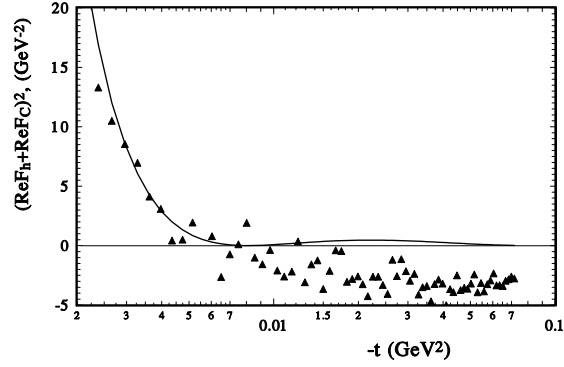}
\end{center}
\vspace{1.cm}
\caption{ The value $\Delta_{R}^{th.}$ and experimental $\Delta_{R}^{exp.}$ are calculated
with the parameters determined by TOTEM  Collaboration \cite{T66} with $\rho(t=0)=0.09$.
  }
\label{Fig_3}
\end{figure}
\begin{figure}
%
\begin{center}
\includegraphics[width=0.7\textwidth]{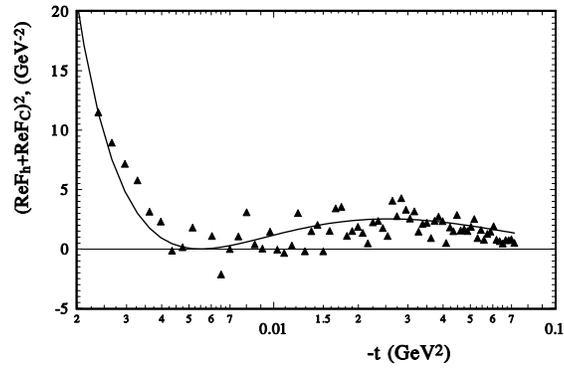}
\end{center}
\vspace{1.cm}
\caption{ The value $\Delta_{R}^{th.}$ and experimental $\Delta_{R}^{exp.}$
 are calculated with the $\rho(t=0) = 0.12$.
  }
\label{Fig_5}
\end{figure}

 Let us determine the value $ \Delta_{R}$ in two ways.
One gives  purely theoretical $\Delta_{R}^{th}(s,t)$
that is dependent on the size of the real part of the scattering amplitude
\begin{eqnarray}
    \Delta_{R}^{th}(s,t)=(ReF_C(t)+ReF_h(s,t))^2.
    \label{Dr}
\end{eqnarray}
  Obviously, it gives the minimum at one point of $t_{min}$ where the Coulomb amplitude equals by module the real part
  of the scattering amplitude.

   Other determination gives the  $\Delta_{R}^{exp}(s,t_{i})$ that is dependent on the experimental data of the differential
   cross sections and the size of the imaginary part of the scattering amplitude.
 \begin{eqnarray}
  \Delta_{R}^{exp}(s,t_{i}) = [\frac{d\sigma}{dt_{i}}|_{exp.} - k \pi *(ImF_c(t_{i}) + ImF_h(t_{i}))^2]/(k \pi).
\label{Drds}
  \end{eqnarray}
  If the real and imaginary parts of the scattering amplitude are indeed determined, then both above values have to be the same.

   Now let us calculate these values using the parameters obtained by the TOTEM Collaboration through the fitting procedure
   of the experimental data at $\sqrt{s}=13$ TeV.
   The results are shown in Fig.3.
   It can be seen that  most parts of the values of $\Delta_{R}^{exp}(s,t_{i})$ have the negative sign.
   It shows the wrong determination of the imaginary part of the scattering amplitude
   or the wrong determination of the normalization of the experimental data.

   A different situation is presented in Fig.4. In this case, the value $\rho(s,t=0) = 0.12$
   and slightly changes the slope of the scattering amplitude.
   The dependence of $\Delta_{R}^{exp.}(s,t_{i})$  and  $ \Delta_{R}^{th}(s,t) $ on momentum transfer is related with
    eqs.(\ref{Dr},\ref{Drds}). Such different results, presented in Fig.3 and Fig.4 show that the real part of the scattering
   amplitude has to be more than $0.1$.

\section{Conclusion}

 Thus, the study of the behavior of the differential cross sections
 in the range of small
 momentum transfer can give  essential information on the behavior of
 the interaction potential at large distances.

     Using only statistical errors and fixing  additional normalization
      of differential cross sections equal to unity,
     we have limited the possible forms of the theoretical representation of the scattering
     amplitude.
     The phenomenological model, the HEGSh model, was used
     for examining the whole region of the momentum transfer
     of two sets of experimental  data obtained by the TOTEM Collaboration at $13$ TeV.
      The simple exponential form of the scattering amplitude was used to examine only
     a small region of  momentum transfer.
     In both cases, an additional fast decreasing term of
     the scattering amplitude was required for a quantitative description of the
     new experimental data.
      The large slope of this term can be connected with a large radius of the hadronic
      interaction and, hence, can be determined by the interaction potential at large distances.
      It can be some part of the hadronic potential responsible for the oscillation behavior
      of the elastic scattering amplitude    \cite{selh95}   

      The discovery of  such anomaly in the behaviour of the differential
      cross section at very small momentum transfer
      in  LHC experiments will give us  important information about
      the behavior of the hadron interaction potential at large distances.
      It may be tightly connected with the problem of  confinement.
      We have shown the existence of such anomaly
      at the statistical  level and  that some other models also
      revealed such unusual behaviour of the scattering amplitude.
      Very likely,  such effects exist also in  experimental data
      at essentially smaller energies \cite{selh95,osc-conf}.
      However, the results of the TOTEM Collaboration have a unique
      unprecedentedly small statistical error and reach minimally  small
      angles of  scattering with the largest number of  experimental points
      in this small region of the momentum transfer.
     The new effects can impact  the determination of
     the sizes of the total cross sections, the ratio of
    the elastic to the total cross sections and the size of the
    $\rho(s,t)$, the ratio of the real to imaginary part of the elastic scattering
    amplitude.
      It is to be noted that the detected new phenomena
      can impact   the determination of the size of the
    $\rho(s,t)$,  the ratio of the real to imaginary part of the elastic scattering
    amplitude and
     the sizes
      of the total cross sections, the ratio of
    the elastic to the total cross sections.

    The comparison of the sizes of
     the total cross sections and $\rho(t=0)$
   obtained for the case with additional coefficient normalization $k$ and  the cases with
    an additional fast decreasing term and $k=1.0$,
     show the large difference.
     If in the first case we obtain $\sigma_{tot (TOTEM)} = 106.2 \pm 0.2$  mb
     which is less than  the value  extracted by the TOTEM      Collaboration
    - $\sigma_{tot (TOTEM)} = 110.6 \pm 3.4$  mb in the analysis of only  small momentum transfer region.
    In this case  the size of $\rho(t=0)= 0.142 \pm 0.004$.
  However, in the case with the $k=1.0$, which require  an additional term with a large slope,
    the value of $\sigma_{tot} = 112.6 \pm 0.11 $ mb which exceeds the  $\sigma_{tot}((TOTEM)$,
    and $\rho(t=0)$ practically coincides with the predictions of the COMPETE Collaboration
    \cite{COMPETE}.
   These results show the necessity the complete analysis of  all the sets of the LHC data
   from $\sqrt{s} = 7$  TeV up to $\sqrt{s} = 13$  TeV including the results of both the Collaborations
   (TOTEM and ATLAS).

\section*{Acknowledgements}
{\it The author would like to thank  V.A. Petrov for the invitation
   to take apart in the conference and the members of the organizing committee
   for the best organization of such interesting conference.}





\begin{thebibliography}{99}
\bibitem{T66}   G.~Antchev {\it et al.} [TOTEM Collaboration],
  arXiv:1812.04732 [hep-ex].

\bibitem{T67}   G.~Antchev {\it et al.} [TOTEM Collaboration],
  arXiv:1812.08283 [hep-ex].

\bibitem{akm} G. Auberson, T. Kinoshita, A. Martin, {\it Phys. Rev.}
            {\bf D 3}, 3185 (1971).

   \bibitem{fp} J. Fischer,
        {\it Phys.Rep.} {\bf 76}, 157 (1981).


\bibitem{royt} S.M. Roy , {\it Phys.Lett.} {\bf B 34}, 407 (1971).

\bibitem{Drem}
  I.M. Dremin, Physics1(1), 33 (2019).

\bibitem{PetOkor-18} V. A. Petrov,
 V. A. Okorokov,
 Int. J. Mod. Phys. {\bf A 33}, 1850077 (2018)
DOI: 10.1142/S0217751X1850077X.



\bibitem{Sel-UA42} O.V. Selyugin, {\it Phys. Lett.} {\bf B 333}, 245 (1994).


     \bibitem{CS-PRL09}  J.~R.~Cudell and O.~V.~Selyugin,
  Phys.\ Rev.\ Lett.\  {\bf 102}  032003 (2009).



  \bibitem{Protvino19}  V.V. Ezhela, V.A. Petrov, N.P. Tkachenko, arxiv: 2003.03817 [hep-ph].




 \bibitem{HEGS0} O.V. Selyugin,
  Eur.\ Phys.\ J.\ C {\bf 72}, 2073 (2012).
  doi:10.1140/epjc/s10052-012-2073-3
 \bibitem{HEGS1} O.V. Selyugin,
  Phys.\ Rev.\ D {\bf 91}, no. 11, 113003 (2015).
  Erratum: [Phys.\ Rev.\ D {\bf 92}, no. 9, 099901 (2015)].
  doi:10.1103/PhysRevD.91.113003, 10.1103/PhysRevD.92.099901


  \bibitem{Orava-Sel}  R. Orava, O. V. Selyugin,  arXiv:1804.05201.


\bibitem{Osc-13} O.V. Selyugin, Phys.Lett., {\bf B 797} 134870 (2019).


\bibitem{hud}  D.J. Hudson, STATISTIC, Lectures on Elementary Statistics and Probability, Geneva (1964).




\bibitem{HEGS-min} O.V. Selyugin,
  Nucl.\ Phys.\ A {\bf 959}  116  (2017).
  doi:10.1016/j.nuclphysa.2017.01.002

\bibitem{HEGSh}  O.~V.~Selyugin, J.-R. Cudell, arxiv: 1810.11538 [ hep-ph ],
 in Proceedings Int. Conference "Diffraction and Low-x 2018", Reggio Calabria, Italy, August 26th t- September 1st, 2018.

   \bibitem{selmp1}
     O.V. Selyugin,
%
  Phys.\ Rev.\ D {\bf 60} 074028   (1999).
 doi:10.1103/PhysRevD.60.074028
%



  \bibitem{GPD-PRD14}   Selyugin  O.V.,
        Phys. Rev., {\bf D 89}, 093007  (2014). 


\bibitem{fd-13} O.V. Selyugin,   
   Modern Physics Letters A, {\bf 36}, No. 18 (2021) 2150148.



  \bibitem{Gribov-Sl} A. Anselm and V. Gribov, Phys.\ Lett.\  B{\bf 40}, 487 (1972).

  \bibitem{Khoze-Sl} V.A. Khoze, A.D. Martin and M.G. Ryskin, arxiv:1410.0508.




\bibitem{RevJenk} R. Fiore, L. Jenkovszky, R. Orava, E. Predazzi, A. Prokudin, and O. Selyugin, "Forward Physics at the LHC Elastic Scattering", Int.J.Mod.Phys. A 2009 24, 2551-2599.



\bibitem{Panch18} S. Pacetti, Y. Strivastava, G. Pancheri,
  Phys.Rev. {bf D 99} (2019) 034014. 


\bibitem{selyf}
   O.V. Selyugin – J. Nucl.Phys. (Yad.Phys.) v.55 (1992).




     \bibitem{selh95} O.V. Selyugin ,
           Ukr.J.Phys. {\bf 41},  (1996)   296.

 \bibitem{osc-conf}        O. V. Selyugin, J. -R. Cudell
        arXiv:1011.4177;
      Talk presented at the International Workshop on Diffraction in High-Energy Physics, Otranto (Lecce, Italy), September 10-15, 2010
      Journal ref: AIP Conf.Proc.1350:115-118,2011.


%
\bibitem{COMPETE}
 J.~R.~Cudell {\it et al.} [COMPETE Collaboration],
  Phys.\ Rev.\ Lett.\  {\bf 89} (2002) 201801.
%





%
%

%
%
%
%
%
%
%
%
%








%





\end{thebibliography}


\end{document}